\documentclass[12pt]{article}
\usepackage{amsfonts}
\usepackage{amsmath}
\newcommand{\non}{\nonumber}

\font\manual=manfnt \def\dbend{\lower3.5pt\hbox{\manual\char127}}

\def\IZ{\relax\ifmmode\mathchoice
{\hbox{\cmss Z\kern-.4em Z}}{\hbox{\cmss Z\kern-.4em Z}}
{\lower.9pt\hbox{\cmsss Z\kern-.4em Z}} {\lower1.2pt\hbox{\cmsss
Z\kern-.4em Z}}\else{\cmss Z\kern-.4em Z}\fi}

\def\rt2{\sqrt{2}}
\def\irt2{{1\over\sqrt{2}}}

\def\hat{\widehat}
\def\slashchar#1{\setbox0=\hbox{$#1$}           
   \dimen0=\wd0                                 
   \setbox1=\hbox{/} \dimen1=\wd1               
   \ifdim\dimen0>\dimen1                        
      \rlap{\hbox to \dimen0{\hfil/\hfil}}      
      #1                                        
   \else                                        
      \rlap{\hbox to \dimen1{\hfil$#1$\hfil}}   
      /                                         
   \fi}
\begin{document}
\begin{titlepage}
\begin{flushright}
\end{flushright}
\vskip 1.5in
\begin{center}
{\bf\Large{Supersymmetric Probes in a  Rotating 5D Attractor}}
\vskip 0.5in {Wei Li and Andrew Strominger} \vskip 0.3in {\small{
\textit{ Jefferson Physical Laboratory, Harvard University,
Cambridge MA 02138, USA}}}
\end{center}
\vskip 0.5in

\baselineskip 16pt

\date{}
\begin{abstract}
Supersymmetric zero-brane and one-brane probes in the squashed
$AdS_2\times S^3$ near-horizon geometry of the BMPV black hole are
studied. Supersymmetric zero-brane probes stabilized by orbital
angular momentum on the $S^3$ are found and shown to saturate a
BPS bound. We also find supersymmetric one-brane probes which have
momentum and winding around a $U(1)_L\times U(1)_R$ torus in the
$S^3$ and in some cases are static.

\end{abstract}
\end{titlepage}
\vfill\eject \tableofcontents
\section{Introduction}

The near-horizon attractor geometry of a BPS black hole has twice
as many supersymmetries as the full asymptotically flat solution.
In four dimensions, such geometries admit  BPS probe
configurations which preserve only near-horizon supersymmetries,
and break all of the supersymmetries of the original
asymptotically flat solution \cite{Simons:2004nm}.   A novel
feature of these configurations is that branes and anti-branes
antipodally located on the $S^2$ preserve the same
supersymmetries. Quantization of these classical configurations
leads to lowest Landau levels which tile the black hole horizon
\cite{Gaiotto:2004pc}. In some cases the degeneracies saturate the
Bekenstein-Hawking black hole entropy \cite{Gaiotto:2004ij}.
Furthermore, an appropriate  expansion of the black hole partition
function in a dilute gas of these states \cite{Gaiotto:2006ns}
yields a derivation of the OSV relation \cite{Ooguri:2004zv}.

These interesting 4D phenomena should all have closely related 5D
cousins \cite{Gaiotto:2005gf}.  With this in mind, the present
paper extends the 4D classical BPS probe analysis of
\cite{Simons:2004nm} to five dimensions. The 5D problem is
considerably enriched by the fact that 5D BMPV BPS black holes can
carry angular momentum $J$ and have a $U(1)_L\times SU(2)_R$
rotational isometry group \cite{Breckenridge:1996is}.   BPS
zero-brane probes that orbit the $S^3$ are found using a
$\kappa$-symmetry analysis. Their location in $AdS_2$ depends on
the azimuthal angle on $S^3$, the background rotation $J$, and the
angular momentum of the probe. For one-branes, we find BPS
configurations with momentum and winding around a torus generated
by a $U(1)_L\times U(1)_R$ rotational subgroup.\footnote{Inclusion
of these states in the partition function of \cite{Gaiotto:2006ns}
could lead to non-factorizing corrections to the OSV relation.} A
one-brane in five dimensions can carry the magnetic charge dual to
the electric charge supporting the BMPV black hole. Interestingly,
we find that this allows for static BPS ``black ring''
configurations, where the angular momentum required for saturation
of the BPS bound is carried by the gauge field.

\section{Review of the BMPV black hole}
The 5D $\mathcal{N}=2$ supersymmetric rotating black hole arises
from M2-branes wrapping holomorphic curves of a  Calabi-Yau
threefold $X$. It is characterized by electric charges $q_{A}$,
$A=1,2,..b_2(X)$, and the angular momentum $J$ in
$SU(2)_{\textrm{left}}$. The metric is \cite{Breckenridge:1996is}
\begin{eqnarray}
ds^2&=&-\left(1+\frac{ Q }{r^2}\right)^{-2}\left[dt+\frac{J}{2r^2}
\sigma_3 \right]^2+\left(1+\frac{ Q }{r^2}\right)
\left(dr^2+r^2d\Omega^2_3\right),\\
d\Omega^2_3&=&\frac{1}{4}\left[d\theta^2+d\phi^2+d\psi^2+2\cos{\theta}d\psi
d\phi\right]=\frac{1}{4}\sum^{3}_{i=1}(\sigma_i)^2,
\end{eqnarray}
where the ranges of the angular parameters are
\begin{equation}
\theta\in[0,\pi],\qquad \phi\in[0,2\pi],\qquad \psi\in[0,4\pi].
\end{equation}
$\sigma_i$ are the right-invariant one-forms:\footnote{The $SU(2)$
rotation matrix is parameterized as:
\begin{displaymath}
e^{i\frac{\sigma_z}{2}\psi}e^{i\frac{\sigma_y}{2}\theta}e^{i\frac{\sigma_z}{2}\phi}=
\left(%
\begin{array}{cc}
   \cos{\frac{\theta}{2}}e^{i(\psi+\phi)/2} &  \sin{\frac{\theta}{2}}e^{i(\psi-\phi)/2} \\
-\sin{\frac{\theta}{2}}e^{-i(\psi-\phi)/2} & \cos{\frac{\theta}{2}}e^{-i(\psi+\phi)/2} \\
    \end{array}%
\right).
\end{displaymath}
}
\begin{eqnarray}
\sigma_1&=&-\sin{\psi}d\theta+\cos{\psi}\sin{\theta}d\phi,\non\\
\sigma_2&=&\cos{\psi}d\theta+\sin{\psi}\sin{\theta}d\phi,\\
\sigma_3&=&d\psi+\cos{\theta}d\phi,\non
\end{eqnarray}
and we choose Planck units $l_5=({\frac{4G_5}{\pi}})^{1/3}=1$. The
graviphoton charge Q is determined via the equations
\begin{equation}
Q^{\frac{3}{2}}=D_{ABC}y^Ay^By^C,
\end{equation}
\begin{equation}
q_A=3D_{ABC}y^By^C,
\end{equation}
with $D_{ABC}$ the intersection form on $X$.

The near-horizon limit ($r \rightarrow 0$) of the metric is
\begin{equation}\label{nbn}
ds^2=-\left[\frac{r^2}{ Q }dt+\frac{J }{2Q}\sigma_3\right]^2+ Q
\frac{dr^2}{r^2}+ Q d\Omega^2_3.
\end{equation}
Rescaling $t$ to absorb $ Q $, defining $\sin^2{B}=\frac{J^2}{ Q^3
}$ and $r^2=1/\sigma$, we obtain the metric in Poincar\'{e}
coordinates:
\begin{equation}\label{Poincare}
ds^2=\frac{Q}{4}\left[-(\frac{dt}{\sigma}+\sin{B}\sigma_3)^2+
\frac{d\sigma^2}{\sigma^2}+\sigma^2_1+\sigma^2_2+\sigma^2_3\right].
\end{equation}
The graviphoton field strength in these coordinates is
\begin{equation}\label{PoincareAF}
F_{[2]}=dA_{[1]}, \qquad \qquad
A_{[1]}=\frac{\sqrt{Q}}{2}[\frac{1}{\sigma}dt+\sin{B}\sigma_3].
\end{equation} We will
also be using the metric in the global coordinates
$(\tau,\chi,\theta,\phi,\psi)$:\footnote{The coordinate
transformation between the global coordinates and Poincar\'{e}
ones is:
\begin{eqnarray}\label{CTcADS3S^3}
t&=&\frac{\cos{B}\cosh{\chi} \sin{\tau}}{\cosh{\chi}
\cos{\tau}+\sinh{\chi}}, \non\\
\sigma &=& \frac{1}{\cosh{\chi} \cos{\tau}+\sinh{\chi}}, \non\\
\psi^{\textrm{Poincar\'{e}}} &=&\psi^{\textrm{global}}+ 2\tan{B}
\tanh^{-1}{(e^{-\chi}\tan{\frac{\tau}{2}})}. \non
\end{eqnarray}}
\begin{equation}\label{Global}
ds^2= \frac{Q}{4}\left[-\cosh^2{\chi}d\tau^2+d\chi^2
+\left(\sin{B}\sinh{\chi}d\tau-\cos{B}\sigma_3\right)^2+\sigma^2_1+\sigma^2_2\right],
\end{equation}
in which \begin{equation}\label{globalAF}
A_{[1]}=\frac{\sqrt{Q}}{2}[\cos{B}\sinh{\chi}d\tau+\sin{B}\sigma_3].
\end{equation}

The near horizon geometry of the BMPV black hole is a kind of
squashed $AdS_2 \times S^3$. The near-horizon isometry supergroup
is $SU(1,1|2)\times U(1)_{\textrm{left}}$, where the bosonic
subgroup of $SU(1,1|2)$ is $SU(1,1)\times SU(2)_{\textrm{right}}$
\cite{Gauntlett:1998fz}. When $J=0$, $U(1)_{\textrm{left}}$ is
promoted to $SU(2)_{\textrm{left}}$ and the full $SO(4)\cong
SU(2)_{\textrm{right}} \times SU(2)_{\textrm{left}}$ rotational
invariance is restored. The unbroken rotational symmetries for
$J\neq 0$ are generated by the Killing vectors
\begin{eqnarray}
\xi^{L}_3&=&\partial_{\psi}
\end{eqnarray}
and
\begin{eqnarray}
\xi^{R}_1&=&\sin{\phi}\partial_{\theta}+\cos{\phi}(\cot{\theta}\partial_{\phi}-\csc{\theta}\partial_{\psi}),\non\\
\xi^{R}_2&=&\cos{\phi}\partial_{\theta}-\sin{\phi}(\cot{\theta}\partial_{\phi}-\csc{\theta}\partial_{\psi}),\\
\xi^{R}_3&=&\partial_{\phi}.\non
\end{eqnarray}

The supersymmetries arise from Killing spinors $\epsilon$ which
are the solutions of the equation
\begin{equation}
\left[d+\frac{1}{4}\omega_{ab}\Gamma^{ab}+\frac{i}{8}\left(e^a
\Gamma^{bc}\Gamma_{a}
F_{bc}-4e^a\Gamma^bF_{ab}\right)\right]\epsilon=0
\end{equation}
To solve this in global coordinates we choose the vielbein
\begin{eqnarray}\label{vielbeinGlobal}
e^0&=&\frac{\sqrt{Q}}{2}[\cosh{(\sin{B}\cos{B}\psi)}\cosh{\chi}d\tau+\sinh{(\sin{B}\cos{B}\psi)}d\chi],\non\\
e^1&=&\frac{\sqrt{Q}}{2}[\sinh{(\sin{B}\cos{B}\psi)}\cosh{\chi}d\tau+\cosh{(\sin{B}\cos{B}\psi)}d\chi],\non\\
e^2&=&\frac{\sqrt{Q}}{2}[-\sin{(\cos^2{B}\psi)}d\theta+\cos{(\cos^2{B}\psi)}\sin{\theta}d\phi],\\
e^3&=&\frac{\sqrt{Q}}{2}[\cos{(\cos^2{B}\psi)}d\theta+\sin{(\cos^2{B}\psi)}\sin{\theta}d\phi],\non\\
e^4&=&\frac{\sqrt{Q}}{2}[-\sin{B}\sinh{\chi}d\tau+\cos{B}\sigma_3].\non
\end{eqnarray}
The Killing spinors are then
\cite{Alonso-Alberca:2002gh}\cite{Alonso-Alberca:2002wr}
\begin{eqnarray}\label{KSGlobal}
\epsilon&=&e^{\left[-\frac{1}{2}(\sin{B}\cos{B}\Gamma^{01}+\cos^2{B}\Gamma^{23})\psi\right]}e^{\left[+\frac{1}{2}(\cos{B}\Gamma^{24}+i\sin{B}\Gamma^{2})\theta\right]}e^{\left[-\frac{1}{2}(\cos{B}\Gamma^{34}+i\sin{B}\Gamma^{3})\phi\right]}\non\\
&&e^{\left[+\frac{1}{2}(\sin{B}\Gamma^{04}-i\cos{B}\Gamma^{0})\chi\right]}e^{\left[-\frac{1}{2}(\sin{B}\Gamma^{14}-i\cos{B}\Gamma^{1})\tau\right]}\epsilon_0\non\\
& \equiv& S \epsilon_0,
\end{eqnarray}
where $\epsilon_0$ is any spinor with constant components in the
frame (\ref{vielbeinGlobal}).

For Poincar\'{e} coordinates we  choose the vielbein
\begin{equation}\label{vielbeinPoincare}
e^0=\frac{\sqrt{Q}}{2}[\frac{dt}{\sigma}+\sin{B}\sigma_3], \qquad
e^1=\frac{\sqrt{Q}}{2}\frac{d\sigma}{\sigma}, \qquad
e^2=\frac{\sqrt{Q}}{2}\sigma_1, \qquad
e^3=\frac{\sqrt{Q}}{2}\sigma_2,\qquad
e^4=\frac{\sqrt{Q}}{2}\sigma_3.
\end{equation}
The Killing spinors are \cite{Gauntlett:1998fz}
\begin{eqnarray}\label{KSPoincare}
\epsilon^{+}&=&\frac{1}{\sqrt{\sigma}}R(\theta, \phi, \psi)\epsilon^{+}_0,\\
\epsilon^{-}&=&\left[\sqrt{\sigma}(1-\sin{B}\Gamma^{04})-\frac{t}{\sqrt{\sigma}}\Gamma^{01}\right]R(\theta,
\phi, \psi)\epsilon^{-}_0,
\end{eqnarray}
where
\begin{eqnarray}
R(\theta, \phi,
\psi)&=&e^{-\frac{1}{2}\Gamma^{23}\psi}e^{\frac{1}{2}\Gamma^{24}\theta}e^{-\frac{1}{2}\Gamma^{23}\phi},\non\\
i\Gamma^0\epsilon^{\pm}_0&=&\pm\epsilon^{\pm}_0,
\end{eqnarray}
for constant $\epsilon^\pm_0$.
\section{Supersymmetric probe configurations}
In this section, we find classical brane trajectories which
preserve some supersymmetries of the rotating attractor
(\ref{nbn}).  The worldvolume action has a local $\kappa$-symmetry
(parameterized by $\kappa$) as well as a spacetime supersymmetry
transformation (parameterized by $\epsilon$) which acts
nonlinearly. A spacetime supersymmetry is preserved if its action
on the worldvolume fermions $\Theta$ can be compensated by a
$\kappa$ transformation
\cite{Becker:1995kb}\cite{Bergshoeff:1997kr}:
\begin{equation}
\delta_{\epsilon}\Theta+\delta_{\kappa}
\Theta=\epsilon+(1+\Gamma)\kappa(\sigma)=0,
\end{equation}
where $\Gamma$ is given in various cases analyzed below. This
gives the condition
\begin{equation}\label{kappa}
(1-\Gamma)\epsilon=0,
\end{equation}
which must be solved for both the Killing spinor and the probe
trajectory.

\subsection{Zero-brane probe}

For the zero-brane the (bosonic part of the) $\kappa$-symmetry
projection operator is
\begin{equation}
\Gamma=\frac{1}{\sqrt{h_{00}}}\tilde{\Gamma}_{0},
\end{equation}
where $h$ and $\tilde{\Gamma}_{0}$ are the pull-backs of the
metric and Dirac matrix onto the worldline of the zero-brane,
respectively:
\begin{eqnarray}
h_{00}&=&\partial_{0}X^{\mu}\partial_{0}X^{\nu}G_{\mu\nu},\\
\tilde{\Gamma}_{0}&=&\partial_{0}X^{\mu}e^a_{\mu}\Gamma_a.
\end{eqnarray}

\subsubsection{Global coordinates}\label{kappasymmetryGlobal}

First, let's look at the global coordinates. In the static gauge,
where we set the worldvolume time $\sigma^0$ equal to the global
time $\tau$, the $\kappa$-symmetry operator is
\begin{equation}
\Gamma=\frac{1}{\sqrt{h_{00}}}\frac{dX^{\mu}}{d\tau}e^a_{\mu}\Gamma_a.
\end{equation}

To solve for the classical trajectory of a supersymmetric
zero-brane, we plug the Killing spinors (\ref{KSGlobal}) into the
$\kappa$-symmetry condition (\ref{kappa}) of the supersymmetric
zero-brane. A zero-brane following a classical trajectory, given
by $(\chi(\tau),\theta(\tau),\phi(\tau),\psi(\tau))$, is
supersymmetric if, in the notation of (\ref{KSGlobal}),
\begin{equation}
\frac{1}{\sqrt{h_{00}}} \frac{dX^{\mu}}{d\tau}e^a_{\mu} S^{-1}
\Gamma_a S \epsilon_{0}=\epsilon_{0},
\end{equation}
for some  constant $\epsilon_{0}$, where
$S=S(\chi,\tau,\theta,\phi,\psi)$. The explicit prefactors are
\begin{eqnarray}\label{long}
S^{-1}  e^a_{0}\Gamma_a   S&=&\frac{\sqrt{Q}}{2}[(\cosh{\chi}\cos{\tau}\cos^2{B}+\sin{\theta}\cos{\phi}\sin^2{B})\Gamma^0\non\\
&&+i\cosh{\chi}\sin{\tau}\cos{B}\Gamma^{01}-i\cos{\theta}\sin{B}\Gamma^{02}-i\sin{\theta}\sin{\phi}\sin{B}\Gamma^{03}\non\\
&&+i(\cosh{\chi}\cos{\tau}-\sin{\theta}\cos{\phi})\sin{B}\cos{B}\Gamma^{04}],\non\\
S^{-1}  e^a_{1}\Gamma_a   S&=&(-1)\frac{\sqrt{Q}}{2}[\sin{\theta}\cos{\phi}\cos{\tau}\Gamma^1\non\\
&&-\sin{\tau}\sin{B}e^{\frac{1}{2}(\cos{B}\Gamma^{34}+i\sin{B}\Gamma^{3})\phi}e^{-(\cos{B}\Gamma^{24}+i\sin{B}\Gamma^{2})\theta}e^{\frac{1}{2}(\cos{B}\Gamma^{34}+i\sin{B}\Gamma^{3})\phi}\Gamma^{4}\non\\
&&-i\cos{\tau}\cos{\theta}\sin{B}\Gamma^{12}-i\cos{\tau}\sin{\theta}\sin{\phi}\sin{B}\Gamma^{13}+ie^{(\sin{B}\Gamma^{14}-i\cos{B}\Gamma^{1})\tau}\sinh{\chi}\cos{B}\Gamma^{01}\non\\
&&+i(\cosh{\chi}-\sin{\theta}\cos{\phi}\cos{\tau})\cos{B}(\sin{B}\Gamma^{14}-i\cos{B}\Gamma^{1})],\non\\
S^{-1}  e^a_{2}\Gamma_a   S&=&(-1)\frac{\sqrt{Q}}{2}[\cosh{\chi}\cos{\tau}\cos{\phi}\Gamma^3-\cosh{\chi}\cos{\tau}\sin{\phi}\cos{B}\Gamma^{4}\\
&&+e^{(\cos{B}\Gamma^{34}+i\sin{B}\Gamma^{3})\phi}(+i\sinh{\chi}\cos{B}\Gamma^{03}-i\cosh{\chi}\sin{\tau}\cos{B}\Gamma^{13}+i\cos{\theta}\sin{B}\Gamma^{23})\non\\
&&+i(\cosh{\chi}\cos{\tau}\cos{\phi}-\sin{\theta})\sin{B}(\cos{B}\Gamma^{34}+i\sin{B}\Gamma^{3})],\non\\
S^{-1}  e^a_{3}\Gamma_a   S&=&(-1)\frac{\sqrt{Q}}{2}[(\cosh{\chi}\cos{\tau}\cos^2{B}+\sin{\theta}\cos{\phi}\sin^2{B})\Gamma^2\non\\
&&+i\sinh{\chi}\cos{B}\Gamma^{02}-i\cosh{\chi}\sin{\tau}\cos{B}\Gamma^{12}-i\sin{\theta}\sin{\phi}\sin{B}\Gamma^{23}\non\\
&&+i(\cosh{\chi}\cos{\tau}-\sin{\theta}\cos{\phi})\sin{B}\cos{B}\Gamma^{24}],\non\\
S^{-1}  e^a_{4}\Gamma_a   S&=&(-1)\frac{\sqrt{Q}}{2}\cos{B}e^{+\frac{1}{2}(\sin{B}\Gamma^{14}-i\cos{B}\Gamma^{1})\tau}e^{-(\sin{B}\Gamma^{04}-i\cos{B}\Gamma^{0})\chi}e^{+\frac{1}{2}(\sin{B}\Gamma^{14}-i\cos{B}\Gamma^{1})\tau}\non\\
&&e^{+\frac{1}{2}(\cos{B}\Gamma^{34}+i\sin{B}\Gamma^{3})\phi}e^{-(\cos{B}\Gamma^{24}+i\sin{B}\Gamma^{2})\theta}e^{+\frac{1}{2}(\cos{B}\Gamma^{34}+i\sin{B}\Gamma^{3})\phi}\Gamma^{4}.\non
\end{eqnarray}

We first see  that a probe static in the global time $\tau$ cannot
be supersymmetric. For such a  probe we have
$\frac{d\chi}{d\tau}=\frac{d\theta}{d\tau}=\frac{d\phi}{d\tau}=\frac{d\psi}{d\tau}=0$
and the $\kappa$-symmetry condition reduces to
\begin{eqnarray}\label{static0-brane}
&&\frac{1}{\sqrt{-1-\cos^2{B}\sinh^2{\chi}}}\cdot[(\cosh{\chi}\cos{\tau}\cos^2{B}+\sin{\theta}\cos{\phi}\sin^2{B})\Gamma^0\non\\
&&+i\cosh{\chi}\sin{\tau}\cos{B}\Gamma^{01}-i\cos{\theta}\sin{B}\Gamma^{02}-i\sin{\theta}\sin{\phi}\sin{B}\Gamma^{03}\\
&&+i(\cosh{\chi}\cos{\tau}-\sin{\theta}\cos{\phi})\sin{B}\cos{B}\Gamma^{04}]\epsilon_{0}=\epsilon_{0}.\non
\end{eqnarray}
The terms in this equation proportional to $\cos{\tau}$,
$\sin{\tau}$ and $1$ must all vanish separately, which is clearly
impossible. The lack of such configurations is not surprising,
because angular momentum must be nonzero for a nontrivial BPS
configuration.

Now we allow the probe to orbit around the $S^3$. Solving the
$\kappa$-symmetry condition (\ref{kappa})  using (\ref{long}) for
Killing spinors obeying
\begin{equation}\label{jjh}
\Gamma^{02}\epsilon_{0}=\mp\epsilon_{0},
\end{equation}
we find the supersymmetric trajectory at a generic
$(\chi,\theta,\psi)$ to be
\begin{equation}\label{globalphi}
\frac{d\chi}{d\tau}=\frac{d\theta}{d\tau}=\frac{d\psi}{d\tau}=0,
\qquad \frac{d\phi}{d\tau}=\pm 1.
\end{equation}
This  is a probe orbiting along the $\phi$-direction.

The constraint on the Killing spinor (\ref{jjh}) projects out half
of the components of $\epsilon_{0}$, i.e. the orbiting zero-brane
probe is a half-BPS configuration. We will show in the next
subsection, using the BPS bound, that this supersymmetric
trajectory is unique up to rotations.

\subsubsection{A BPS bound}

The worldline action of a zero brane probe, with mass $m$ and the
electric charge $q$, can be written as
\begin{equation}
S=-m \int \sqrt{h} d\sigma^0 +q\int A_{[1]},
\end{equation}
where $A_{[1]}$ is the 1-form gauge field (\ref{globalAF}). We
consider supersymmetric probes which have $q=m$.\footnote{The
zero-brane can be obtained by wrapping M2-branes on the
holomorphic two-cycles of the Calabi-Yau threefold $X$. It carries
electric charges $v_{A}$, $A=1,2,..b_2(X)$. Then
$m=q=\frac{v_{A}y^{A}}{\sqrt{Q}/2}$.}

In global coordinates with $\sigma^0=\tau$, the Lagrangian of the
system is
\begin{eqnarray}
\mathcal{L}&=&\frac{\sqrt{Q}}{2}\{-m\sqrt{\cosh^2{\chi}-\dot{\chi}^2-[\sin{B}\sinh{\chi}-\cos{B}(\dot{\psi}+\cos{\theta}\dot{\phi})]^2-\dot{\theta}^2-\sin^2{\theta}\dot{\phi}^2}\non\\
&&+m[\cos{B}\sinh{\chi}+\sin{B}(\dot{\psi}+\cos{\theta}\dot{\phi})]\}.
\end{eqnarray}
The corresponding Hamiltonian is
\begin{eqnarray}
H&=&\cosh{\chi}\sqrt{P_{\chi}^2+P_{\theta}^2+(\frac{\cos{\theta}P_{\phi}-P_{\psi}}{\sin{\theta}})^2+P^2_{\phi}+(\frac{\sin{B}P_{\psi}-\frac{\sqrt{Q}}{2}m}{\cos{B}})^2}+\sinh{\chi}(\frac{\sin{B}P_{\psi}-\frac{\sqrt{Q}}{2}m}{\cos{B}}),\non
\end{eqnarray}
where the momenta are
\begin{eqnarray}
P_{\chi}&=& \frac{m\sqrt{Q}}{2\sqrt{
h}}{\dot{\chi}},\non\\
P_{\theta}&=& \frac{m\sqrt{Q}}{2\sqrt{
h}}{\dot{\theta}},\\
P_{\phi}&=& \frac{m\sqrt{Q}}{2}\left[\frac{1}{\sqrt{h}}\left(-\cos{B}\cos{\theta}[\sin{B}\sinh{\chi}-\cos{B}(\dot{\psi}+\cos{\theta}\dot{\phi})]+\sin^2{\theta}\dot{\phi}\right)+\sin{B}\cos{\theta}\right],\non\\
P_{\psi}&=&
\frac{m\sqrt{Q}}{2}\left[\frac{1}{\sqrt{h}}\left(-\cos{B}[\sin{B}\sinh{\chi}-\cos{B}(\dot{\psi}+\cos{\theta}\dot{\phi})]\right)+\sin{B}\right],\non
\end{eqnarray}
and
\begin{equation}
h=\cosh^2{\chi}-\dot{\chi}^2-[\sin{B}\sinh{\chi}-\cos{B}(\dot{\psi}+\cos{\theta}\dot{\phi})]^2-\dot{\theta}^2-\sin^2{\theta}\dot{\phi}^2.
\end{equation}

The unbroken rotational symmetries lead to the conserved charges:
\begin{eqnarray}
J^1_{\textrm{right}}&=&\sin{\phi}P_{\theta}+\cos{\phi}(\cot{\theta}P_{\phi}-\csc{\theta}P_{\psi}),\non\\
J^2_{\textrm{right}}&=&\cos{\phi}P_{\theta}-\sin{\phi}(\cot{\theta}P_{\phi}-\csc{\theta}P_{\psi}),\\
J^3_{\textrm{right}}&=&P_{\phi},\non\\
J^3_{\textrm{left}}&=&P_{\psi}.\non
\end{eqnarray}

It is easy to see that there are no static solutions. They would
have to minimize the potential energy according to
\begin{equation}0=
\frac{\partial H}{\partial
\chi}=\frac{\sqrt{Q}}{2}m\cos{B}\cosh{\chi}(\frac{\cos{B}\sinh{\chi}}{\sqrt{\cos^2{B}\sinh^2{\chi}+1}}-1),
\end{equation}
which has no solutions for  finite $\chi$. Physically, the probe
is accelerated to  $\chi=\pm \infty$.

Now we allow the probe to orbit.   Solutions of this type can be
stabilized by the angular potential. The supersymmetric
configuration turns out to be at constant radius in the $AdS_2$,
i.e. $P_{\chi}=0$. The Hamiltonian is minimized with respect to
$\chi$ when
\begin{equation}\label{staticchi}
\tanh{\chi}=-\frac{1}{
\sqrt{P_{\theta}^2+(\frac{\cos{\theta}P_{\phi}-P_{\psi}}{\sin{\theta}})^2+P^2_{\phi}+(\frac{\sin{B}P_{\psi}-\frac{\sqrt{Q}}{2}m}{\cos{B}})^2}}(\frac{\sin{B}P_{\psi}-\frac{\sqrt{Q}}{2}m}{\cos{B}}).
\end{equation}
The value of $H$ at the minimum is
\begin{equation}
H_{min}=\sqrt{P_{\theta}^2+(\frac{\cos{\theta}P_{\phi}-P_{\psi}}{\sin{\theta}})^2+P^2_{\phi}}=
| \vec{J}_{\textrm{right} }|,
\end{equation}
where
$|\vec{J}_{\textrm{right}}|^2=(J^1_{\textrm{right}})^2+(J^2_{\textrm{right}})^2+(J^3_{\textrm{right}})^2$.
This implies the BPS bound
\begin{equation} H \ge |\vec J_{\textrm{right}}|
\end{equation}
for generic $\chi$.

Up to spatial rotations, we may always choose static BPS solutions
to satisfy
\begin{equation}
H=J^3_{\textrm{right}}=\pm
P_\phi,~~~~~~~J^1_{\textrm{right}}=J^2_{\textrm{right}}=0.
\end{equation}
This implies
\begin{equation}\label{angularP}
P_{\theta}=0, \qquad \cos{\theta}P_{\phi}=P_{\psi}.
\end{equation}
Hence, the azimuthal angle is determined by the ratio of left and
right angular momenta:
\begin{equation}
\cos \theta =\frac{J^3_{\textrm{left}}}{J^3_{\textrm{right}}}.
\end{equation}

We can rewrite $\dot{\phi}$ and $\dot{\psi}$ in terms of
$P_{\phi}$ and $P_{\psi}$. With $\dot{\chi}=\dot{\theta}=0$,
\begin{eqnarray}
\dot{\phi}&=&\frac{\cosh{\chi}(\frac{P_{\phi}-\cos{\theta}P_{\psi}}{\sin^2{\theta}})}{
\sqrt{P_{\theta}^2+(\frac{\cos{\theta}P_{\phi}-P_{\psi}}{\sin{\theta}})^2+P^2_{\phi}+(\frac{\sin{B}P_{\psi}-\frac{\sqrt{Q}}{2}m}{\cos{B}})^2}},\\
\dot{\psi}&=&\frac{\cosh{\chi}[\tan{B}(\frac{\sin{B}P_{\psi}-\frac{\sqrt{Q}}{2}m}{\cos{B}})-(\frac{\cos{\theta}P_{\phi}-P_{\psi}}{\sin^2{\theta}})]}{
\sqrt{P_{\theta}^2+(\frac{\cos{\theta}P_{\phi}-P_{\psi}}{\sin{\theta}})^2+P^2_{\phi}+(\frac{\sin{B}P_{\psi}-\frac{\sqrt{Q}}{2}m}{\cos{B}})^2}}+\tan{B}\sinh{\chi}.
\end{eqnarray}
Eliminate $\chi$ through (\ref{staticchi}),
\begin{eqnarray}
\dot{\phi}&=&\frac{1}{\sqrt{P_{\theta}^2+(\frac{\cos{\theta}P_{\phi}-P_{\psi}}{\sin{\theta}})^2+P^2_{\phi}}}(\frac{P_{\phi}-\cos{\theta}P_{\psi}}{\sin^2{\theta}}),\\
\dot{\psi}&=&\frac{1}{\sqrt{P_{\theta}^2+(\frac{\cos{\theta}P_{\phi}-P_{\psi}}{\sin{\theta}})^2+P^2_{\phi}}}(\frac{P_{\psi}-\cos{\theta}P_{\phi}}{\sin^2{\theta}}).
\end{eqnarray}
Plug in (\ref{angularP}), the solution is
\begin{equation}
\dot{\theta}=0, \qquad \qquad \dot{\phi}=\pm 1,\qquad \qquad
\dot{\psi}=0,
\end{equation}
for which $(P_{\phi},P_{\psi})$ are
\begin{eqnarray}
P_{\psi}&=&\pm\frac{\sqrt{Q}}{2}m\frac{\cos{\theta}}{\cos{B}\sinh{\chi}\pm\sin{B}\cos{\theta}},\\
P_{\phi}&=&\pm
\frac{\sqrt{Q}}{2}m\frac{1}{\cos{B}\sinh{\chi}\pm\sin{B}\cos{\theta}}.
\end{eqnarray}
The energy of the particle following this trajectory is equal to
$\pm P_{\phi}$:
\begin{equation}
H=\frac{\sqrt{Q}}{2}m\frac{1}{\cos{B}\sinh{\chi}\pm\sin{B}\cos{\theta}}=\pm
P_{\phi}.
\end{equation}
We see that the solution with $\dot{\phi}=1$ ($\dot{\phi}=-1$)
corresponds to a chiral (anti-chiral) BPS configuration.

Therefore, we have confirmed that the supersymmetric trajectories
(\ref{globalphi}) obtained by solving the $\kappa$-symmetry
condition correspond to the BPS states.

\subsubsection{Poincar\'{e} coordinates}

In Poincar\'{e} coordinates and static gauge $\sigma^0=t$, the
$\kappa$-symmetry condition for a static probe is
\begin{equation}
\frac{1}{\sqrt{-\frac{1}{\sigma^2}}}\left[-\frac{1}{\sigma}\Gamma^{0}\right]\epsilon=i\Gamma^0\epsilon=\epsilon.
\end{equation}
This equation is solved by simply taking
$\epsilon=\epsilon^{+}=\frac{1}{\sqrt{\sigma}}R(\theta, \phi,
\psi)\epsilon^{+}_0$.  Again, we find a half-supersymmetric
solution, although the broken supersymmetries are different than
in the global case. It can be seen that there are no
supersymmetric orbiting trajectories in Poincar\'{e} time.

\subsection{One-brane probe}

In this subsection, we find some supersymmetric one-brane
configurations. We consider a specific Ansatz with no worldvolume
electromagnetic field and with the one-brane geometry:
\begin{eqnarray}\label{ans}
\tau&=&\sigma^0,\non\\
\phi&=&\dot{\phi}\sigma^0+{\phi'}  \sigma^1,\\
\psi&=&\dot{\psi}\sigma^0+\psi'\sigma^1,\non
\end{eqnarray}
where $(\sigma^0, \sigma^1)$ are worldvolume coordinates, and
$\dot \phi,~\dot\psi,~\phi'$ and $\psi'$ are all taken to be
constant. Note that since $(\psi, \phi)$ are the orbits of
$(J^3_L,J^3_R)$, they may be viewed as one-brane momentum-winding
modes on the torus generated by $(J^3_L,J^3_R)$. This torus
degenerates to a circle at the loci $\theta=\{0,\pi\}$. One-branes
of the form (\ref{ans}) at these loci are therefore static (up to
reparametrizations).

With no electromagnetic field the $\kappa$-symmetry condition
is\footnote{There is a simple kappa-symmetric action in six
dimensions, but not in five.  In 5D we expect an extra scalar
field along with the transverse coordinates to fill out the
supermultiplet. For the case of the M5-brane wrapping a Calabi-Yau
4-cycle, the scalar in the effective one-brane arises as a mode of
the antisymmetric tensor field. The Ansatz of this section
corresponds to taking this extra scalar to be a constant.}
\begin{equation}\label{kappastring}
\frac{1}{2}\epsilon^{ij}\tilde{\Gamma}_{ij}\epsilon=\epsilon,
\end{equation}
where $h$ and $\tilde{\Gamma}_{i}$ are the pull-backs of the 5D
metric and gamma matrices onto the one-brane worldsheet. With the
Ansatz (\ref{ans}),  we have explicitly
\begin{eqnarray}
\tilde{\Gamma}_{0}&=&\Gamma_{\tau}+\dot{\phi}\Gamma_{\phi}+\dot{\psi}\Gamma_{\psi},\\
\tilde{\Gamma}_{1}&=&{\phi'} \Gamma_{\phi}+\psi'\Gamma_{\psi},\\
\frac{1}{2}\epsilon^{ij}\tilde{\Gamma}_{ij}&=&\frac{1}{2\sqrt{{\textrm{det}h}}}[\phi'\Gamma_{\tau\phi}+\psi'
\Gamma_{\tau\psi}+(\dot \phi \psi'-\dot \psi
\phi')\Gamma_{\phi\psi}],
\end{eqnarray}
and
\begin{eqnarray}
h_{00}&=&\frac{Q}{4}\{-\cosh^2{\chi}+[\sin{B}\sinh{\chi}-\cos{B}(\dot{\psi}+\cos{\theta}\dot{\phi})]^2+\sin^2{\theta}~\dot{\phi}^2\},\non\\
h_{11}&=&\frac{Q}{4}\{\cos^2{B}(\psi'+\cos{\theta}{\phi'}  )^2+\sin^2{\theta}~{\phi'}  ^2\},\\
h_{01}&=&\frac{Q}{4}\{[\sin{B}\sinh{\chi}-\cos{B}(\dot{\psi}+\cos{\theta}\dot{\phi})](-\cos{B})(\psi'+\cos{\theta}{\phi'}
)+\sin^2{\theta}~\dot{\phi}{\phi'}  \},\non
\end{eqnarray}
and hence
\begin{equation}
{\textrm{det}h}=(\frac{Q}{4})^2\{\cosh^2{\chi}[\cos^2{B}(\psi'+\cos{\theta}\phi')^2+\sin^2{\theta}{\phi'}^2]-
\sin^2{\theta}[\sin{B}\sinh{\chi}\phi'-\cos{B}(-\psi'\dot{\phi}+\phi'\dot{\psi})]^2\}.
\end{equation}

It is simplest to analyze  the $\kappa$-symmetry condition in the form
\begin{equation}\label{kappastringe_0}
S^{-1} \frac{1}{2}\epsilon^{ij}\tilde{\Gamma}_{ij}
S\epsilon_{0}=\epsilon_{0}.
\end{equation}
The rotated gamma matrices appearing in this expression are explicitly
\begin{eqnarray}
&&S^{-1}\Gamma_{\tau\phi} S
\\
&=&-\frac{Q}{4}[(\cosh^2{\chi}\cos^2{B}+\sin^2{\theta}\sin^2{B})\Gamma^{02}-i(\cosh{\chi}\cos{\tau}\cos^2{B}+\sin{\theta}\cos{\phi}\sin^2{B}\non\\
&&-i\cosh{\chi}\sin{\tau}\cos{B}\Gamma^{1}+i\sin{\theta}\sin{\phi}\sin{B}\Gamma^{3}\non\\
&&-i(\cosh{\chi}\cos{\tau}-\sin{\theta}\cos{\phi})\sin{B}\cos{B}\Gamma^{4})(\cos{\theta}\sin{B}\Gamma^{0}+\sinh{\chi}\cos{B}\Gamma^{2})],\non
\end{eqnarray}
\begin{eqnarray}
&&S^{-1}\Gamma_{\tau \psi} S
\\
&=&\frac{Q}{4}\cos{B}\{-\cosh^2{\chi}\cos{\theta}\cos{B}\Gamma^{02}\non\\
&&+\cos{B}\sinh{\chi}[i\cosh{\chi}\sin{\theta}\cos{\tau}\cos{\phi}\Gamma^{4}+\cosh{\chi}\sin{\theta}\sin{\phi}\sin{\tau}\Gamma^{13}\non\\
&&-\cosh{\chi}\sin{\theta}\cos{\tau}\sin{\phi}(\sin{B}\Gamma^{34}-i\cos{B}\Gamma^{3})\non\\
&&+\cosh{\chi}\sin{\theta}\cos{\phi}\sin{\tau}(\cos{B}\Gamma^{14}+i\sin{B}\Gamma^{1})]\non\\
&&-(\cos^2{B}\cosh^2{\chi}\sin{\theta}\cos{\phi}+\sin^2{B}\cosh{\chi}\cos{\tau})\Gamma^{04}-\sin{B}\cos{B}\cosh{\chi}\sinh{\chi}\cos{\tau}\cos{\theta}\Gamma^{24}\non\\
&&-\cosh{\chi}\sin{\tau}\sin{B}\Gamma^{01}+\cos{B}\cosh{\chi}\sinh{\chi}\cos{\theta}\sin{\tau}\Gamma^{12}\non\\
&&-\cosh^2{\chi}\sin{\theta}\sin{\phi}\cos{B}\Gamma^{03}\non\\
&&-i\cosh{\chi}(\cosh{\chi}\sin{\theta}\cos{\phi}-\cos{\tau})\sin{B}\cos{B}\Gamma^{0}+i\cos^2{B}\cosh{\chi}\sinh{\chi}\cos{\tau}\cos{\theta}\Gamma^{2}\},\non
\end{eqnarray}
\begin{eqnarray}
&&
S^{-1}\Gamma_{\phi\psi} S \\
&=&\frac{Q}{4}\cos{B}\{+\sinh{\chi}\sin^2{\theta}\sin{B}\Gamma^{02}\non\\
&&+\sin{B}\cos{\theta}[i\cosh{\chi}\sin{\theta}\cos{\phi}\cos{\tau}\Gamma^{4}+\cosh{\chi}\sin{\theta}\sin{\phi}\sin{\tau}\Gamma^{13}\non\\
&&+\cosh{\chi}\sin{\theta}\cos{\phi}\sin{\tau}(\cos{B}\Gamma^{14}+i\sin{B}\Gamma^{1})\non\\
&&-\cosh{\chi}\sin{\theta}\sin{\phi}\cos{\tau}(\sin{B}\Gamma^{34}-i\cos{B}\Gamma^{3})]\non\\
&&-\sin{B}\cos{B}\sinh{\chi}\sin{\theta}\cos{\theta}\cos{\phi}\Gamma^{04}+(\cosh{\chi}\sin^2{\theta}\cos{\tau}\sin^2{B}+\sin{\theta}\cos{\phi}\cos^2{B})\Gamma^{24}\non\\
&&-\cosh{\chi}\sin^2{\theta}\sin{\tau}\sin{B}\Gamma^{12}\non\\
&&-\sin{B}\sin{\theta}\cos{\theta}\sin{\phi}\sinh{\chi}\Gamma^{03}+\sin{\theta}\sin{\phi}\cos{B}\Gamma^{23}\non\\
&&-i\sin^2{B}\sinh{\chi}\sin{\theta}\cos{\theta}\cos{\phi}\Gamma^{0}-i\sin{\theta}(\cosh{\chi}\sin{\theta}\cos{\tau}-\cos{\phi})\cos{B}\sin{B}\Gamma^{2}\}.\non
\end{eqnarray}
This all simplifies at points obeying
 \begin{equation}
\sinh{\chi}=\pm\tan{B}\cos{\theta}
\end{equation}
when $-\psi'\dot{\phi}+\phi'\dot{\psi}=\pm \psi'$.  Under these
conditions
\begin{equation}
\sqrt{\textrm{det}h}= \frac{Q}{4}(\phi'+\cos{\theta}\psi'),
\end{equation}
and \begin{eqnarray}\label{rer}
&&S^{-1}[\phi'\Gamma_{\tau\phi}+\psi' \Gamma_{\tau\psi}+(\dot \phi
\psi'-\dot \psi \phi')\Gamma_{\phi\psi}] S\non\\
&=&
\frac{Q}{4}[-(\phi'+\cos{\theta}\psi')\Gamma^{02}+(\phi'\hat{D}_1+\psi'\hat{D}_2)(\Gamma^{0}\pm\Gamma^{2})],
\end{eqnarray}
where
\begin{eqnarray}
\hat{D}_1&=&i\cos{\theta}\sin{B}[\cosh{\chi}\cos{\tau}\cos^2{B}+\sin{\theta}\cos{\phi}\sin^2{B}\non\\
&&-i\cosh{\chi}\sin{\tau}\cos{B}\Gamma^{1}+i\sin{\theta}\sin{\phi}\sin{B}\Gamma^{3}-i(\cosh{\chi}\cos{\tau}-\sin{\theta}\cos{\phi})\sin{B}\cos{B}\Gamma^{4}],\non\\
\hat{D}_2&=&-\cos{B}(\cos^2{B}\sin{\theta}\cos{\phi}+\sin^2{B}\cosh{\chi}\cos{\tau})\Gamma^{4}+\cos{B}\sin{B}\cosh{\chi}\sin{\tau}\Gamma^{1}\non\\
&&+\cos^2{B}\sin{\theta}\sin{\phi}\Gamma^{3}-i\sin{B}\cos^2{B}(\sin{\theta}\cos{\phi}-\cosh{\chi}\cos{\tau}).\non
\end{eqnarray}
So far we have not chosen which supersymmetries are to be
preserved. We take those generated by spinors obeying
$\Gamma^{02}\epsilon_0=\pm\epsilon_0$, or equivalently
$\Gamma^{2}\epsilon_0=\mp\Gamma^0 \epsilon_0$. In this case, the
last term in (\ref{rer}) can be dropped and the supersymmetry
conditions are satisfied.

To summarize,  any configuration satisfying
\begin{eqnarray}
-\psi'\dot{\phi}+\phi'\dot{\psi}&=&\pm \psi', \qquad \dot{\chi}=\dot{\theta}=0,\non\\
\sinh{\chi}&=&\pm\tan{B}\cos{\theta}
\end{eqnarray}
preserves those supersymmetries corresponding to
\begin{equation}\label{ttt}
\Gamma^{02}\epsilon_0=\pm\epsilon_0.
\end{equation}
Other BPS configurations preserving other sets of supersymmetries
can be obtained by $SL(2,R)\times SO(4)$ rotations of these ones.

Note that, as for the zero-branes,  there are generic solutions
for any $\theta$. These include $\theta=\{0,\pi\}$, which
correspond to static one-branes because the $(\psi,\phi)$ torus
degenerates to a circle along these loci. Static solutions are
possible because a one-brane probe in 5D couples magnetically to
the dual of the spacetime gauge field $F_{[2]}$ of
(\ref{globalAF}) hence there is nonzero angular momentum carried
by the fields.

\section{Acknowledgments}
We would like to thank M. Ernebjerg, D. Gaiotto, L. Huang, J.
Lapan and X. Yin for many helpful discussions. The work was
supported by DOE grant DE-FG02-91ER40654.


\begin{thebibliography}{99}

\bibitem{Simons:2004nm}
  A.~Simons, A.~Strominger, D.~M.~Thompson and X.~Yin,
  ``Supersymmetric branes in AdS(2) x S**2 x CY(3),''
  Phys.\ Rev.\ D {\bf 71}, 066008 (2005)
  [arXiv:hep-th/0406121].

\bibitem{Gaiotto:2004pc}
  D.~Gaiotto, A.~Simons, A.~Strominger and X.~Yin,
  ``D0-branes in black hole attractors,''
  arXiv:hep-th/0412179.

\bibitem{Gaiotto:2004ij}
  D.~Gaiotto, A.~Strominger and X.~Yin,
  ``Superconformal black hole quantum mechanics,''
  arXiv:hep-th/0412322.

\bibitem{Gaiotto:2006ns}
  D.~Gaiotto, A.~Strominger and X.~Yin,
  arXiv:hep-th/0602046.

\bibitem{Ooguri:2004zv}
  H.~Ooguri, A.~Strominger and C.~Vafa,
  Phys.\ Rev.\ D {\bf 70}, 106007 (2004)
  [arXiv:hep-th/0405146].

\bibitem{Gaiotto:2005gf}
  D.~Gaiotto, A.~Strominger and X.~Yin,
  ``New connections between 4D and 5D black holes,''
  JHEP {\bf 0602}, 024 (2006)
  [arXiv:hep-th/0503217].

\bibitem{Breckenridge:1996is}
  J.~C.~Breckenridge, R.~C.~Myers, A.~W.~Peet and C.~Vafa,
  ``D-branes and spinning black holes,''
  Phys.\ Lett.\ B {\bf 391}, 93 (1997)
  [arXiv:hep-th/9602065].

\bibitem{Alonso-Alberca:2002gh}
  N.~Alonso-Alberca, E.~Lozano-Tellechea and T.~Ortin,
  ``Geometric construction of Killing spinors and supersymmetry algebras in
  homogeneous spacetimes,''
  Class.\ Quant.\ Grav.\  {\bf 19}, 6009 (2002)
  [arXiv:hep-th/0208158].

\bibitem{Alonso-Alberca:2002wr}
  N.~Alonso-Alberca, E.~Lozano-Tellechea and T.~Ortin,
  Class.\ Quant.\ Grav.\  {\bf 20}, 423 (2003)
  [arXiv:hep-th/0209069].

\bibitem{Gauntlett:1998fz}
  J.~P.~Gauntlett, R.~C.~Myers and P.~K.~Townsend,
  ``Black holes of D = 5 supergravity,''
  Class.\ Quant.\ Grav.\  {\bf 16}, 1 (1999)
  [arXiv:hep-th/9810204].


\bibitem{Becker:1995kb}
  K.~Becker, M.~Becker and A.~Strominger,
  ``Five-branes, membranes and nonperturbative string theory,''
  Nucl.\ Phys.\ B {\bf 456}, 130 (1995)
  [arXiv:hep-th/9507158].




\bibitem{Bergshoeff:1997kr}
  E.~Bergshoeff, R.~Kallosh, T.~Ortin and G.~Papadopoulos,
  ``Kappa-symmetry, supersymmetry and intersecting branes,''
  Nucl.\ Phys.\ B {\bf 502}, 149 (1997)
  [arXiv:hep-th/9705040].


\end{thebibliography}
\end{document}